\documentclass[aps,prl,superscriptaddress,twocolumn,showpacs,amsmath,amssymb]{revtex4}
\usepackage{stmaryrd}
\usepackage{times}
\usepackage{graphicx}
\usepackage{dcolumn}
\usepackage{bm}

\begin{document}


\title{Thermal transport and non-equilibrium temperature drop across a magnetic nanostructured interface}
\author{Jia Zhang}
\email[Email:]{Jia.Zhang@exp1.physik.uni-giessen.de}
\affiliation{I. Physikalisches Institut, Justus Liebig University Giessen,
Heinrich-Buff-Ring 16, 35392 Giessen, Germany}
\author{Michael Bachman}
\affiliation{I. Physikalisches Institut, Justus Liebig University Giessen,
Heinrich-Buff-Ring 16, 35392 Giessen, Germany}
\author{Michael Czerner}
\affiliation{I. Physikalisches Institut, Justus Liebig University Giessen,
Heinrich-Buff-Ring 16, 35392 Giessen, Germany}
\author{Christian Heiliger}
\email[Email:]{Christian.Heiliger@physik.uni-giessen.de}
\affiliation{I. Physikalisches Institut, Justus Liebig University Giessen,
Heinrich-Buff-Ring 16, 35392 Giessen, Germany}
\date{\today}

\begin{abstract}

In a number of current experiments in the field of spin-caloritronics a temperature gradient across a nanostructured interface is applied and spin-dependent transport phenomena are observed. However, a lack in the interpretation and knowledge let it unclear how the temperature drop across a magnetic nanostructured interface looks like where both phonons and electrons may contribute to thermal transport. We answer this question for the case of a magnetic tunnel junction (MTJ) where the tunneling magneto Seebeck effect occurs. Nevertheless, our results can be extended to other nanostructured interfaces as well. Using an \textit{ab initio} method we explicitly calculate phonon and electron thermal conductance across Fe/MgO/Fe-MTJs by using Green's function method. Further, by estimating the electron-phonon interaction in the Fe leads we are able to calculate the electron and phonon temperature profile across the Fe/MgO/Fe-MTJ.
Our results show that there is an electron-phonon temperature imbalance at the Fe-MgO interfaces. In consequence, a revision of the interpretation of current experimental measurements might be necessary.

\end{abstract}


\maketitle

In the 1990s, Johson and Silsbee~\cite{Johnson:1987} have pioneered the
theoretical investigation of thermal effects in nano-magnetic systems.
However, it took until 2008 for spin-caloritronics~\cite{Bauer:2010,Bauer:2012} phenomenon to emerge after the discovery of the spin Seebeck effect in NiFe films by Uchida \textit{et. al}~\cite{Uchida:2008,Xiao:2010}.
Since then interesting spin-caloritronic transport phenomena are observed including thermal spin injection~\cite{Breton:2011} and magneto-Peltier effect~\cite{Hatami:2009,Flipse:2012,Flipse:2014}.
Magneto-thermoelectric phenomena, e.g. the tunneling magneto-Seebeck (TMS) effect, which occurs in magnetic tunnel junctions (MTJs) and is the dependence of the charge Seebeck coefficient on the magnetic alignments, is demonstrated both in experiments~\cite{Walter:2011,Liebing:2011,Lin:2012} and first-principles calculations~\cite{Czerner:2011,Christian:2013}. Also thermal spin transfer torque has been theoretically predicted~\cite{Jia:2011,Christian:2014} and experiments are on the way~\cite{leut:2013}. This arise the possibility of MgO-MTJs as a potential key structure for spin-caloritronic applications.

All these phenomena have in common that there is a temperature gradient across an interface or across a very thin layer. However, at present the understanding of the temperature drop and even how heat is transported across a magnetic interface is incomplete or unknown. In this letter we focus on MgO-based-MTJs, but our results can at least qualitatively applied to other systems as well.

To get the correct temperature drop across the barrier in a MTJ the appropriate value of the thermal conductance $\kappa^{MTJ}$ is important. Considering the fact that heat is carried by phonons as well as electrons it is not clear if $\kappa^{MTJ}$ is just the sum of both contributions as in the clasical limit or if we need a new definition. In the analysis of the experiments the temperature across the MgO-MTJs has been simulated by COMSOL~\cite{Walter:2011,Liebing:2011} where $\kappa^{MTJ}$ is typically taken from the thermal conductivity of bulk-MgO or MgO thin film or in between, where the variation of this value is one order of magnitude.

In order to check if this approximation is correct one should get a rough idea about the phonon transport length scale in Fe/MgO/Fe-MTJs. For this purpose, the phonon mean free path $\Lambda$ of bulk MgO can be estimated by~\cite{Chen:1998}
\begin{equation}\label{eqn:mfp}
k_p= 1/3 \  C{\upsilon}{\Lambda}
\end{equation}
where $k_p$ is the thermal conductivity, C is the volumetric specific heat, and $\upsilon$ is the average phonon group velocity. By taking experimental values of these quantities for MgO, which are listed in Table~\ref{tab:table1}, the phonon mean free path $\Lambda$ for bulk MgO is around 6.4 nm. Consequently, $\Lambda$ is larger than the typical MgO barrier thickness $d$ (between 0.6$\sim$2 nm) used in MTJs. This means it is not appropriate to take for the thermal conductivity of the junction the bulk value of MgO or the thin film value because these thin films have a thickness of several hundreds of nanometer~\cite{Lee:1998}. Eventually, the vibrational properties of the interface region and the phonon transport across the whole Fe/MgO/Fe junction has to be calculated explicitly.

Further, thinking about a MTJ it is not obvious what is the role of electrons versus phonons for the thermal transport. In the metallic ferromagnetic leads the thermal transport is dominated by electrons whereas in MgO as an insulator the thermal transport is determined by phonons. Moreover, it is not clear what happens at the interface. Are the electronic and phononic systems in equilibrium, e.g. at the same temperature, at the interface? If not are there any consequences for the interpretation of experimental results?

\begin{table}[b]
\caption{\label{tab:table1}Experimental thermal properties of bulk Fe and MgO.}
\centering
\begin{ruledtabular}
\begin{tabular}{c|cc}
Thermal property at 300 K &\textrm{bcc-Fe}&\textrm{fcc-MgO} \\ \hline
\textrm{Phonon thermal conductivity $k_p$ (W/m/K)}&17.55~\cite{Williams:1981}&49.9~\cite{Anne:2014}\\ \hline
\textrm{Electron thermal conductivity $k_e$ (W/m/K)} & 62.75~\cite{Williams:1981} &-\\ \hline
\textrm{Volumetric specific heat C (10$^6$J/m$^3$/K)}& - & 3.345~\cite{Karki:2000}  \\ \hline
\textrm{Average phonon group velocity (m/s)}       & - & 7000~\cite{Glen:1962}     \\ \hline
\textrm{Phonon Mean free path (nm) using Eq.~(\ref{eqn:mfp})} & - & 6.39 \\ 
\end{tabular}
\end{ruledtabular}
\end{table}

To answer these questions, first we calculate the phonon thermal conductance $\kappa_{p}^{MTJ}$ and second we calculate the electron thermal conductance $\kappa_e^{MTJ}$. Consequently, the overall thermal conductance $\kappa^{MTJ}$ is defined and the temperature profile across the Fe/MgO/Fe-MTJs is calculated by estimating the electron-phonon coupling.

\begin{figure}
{\includegraphics[width=8.4cm]{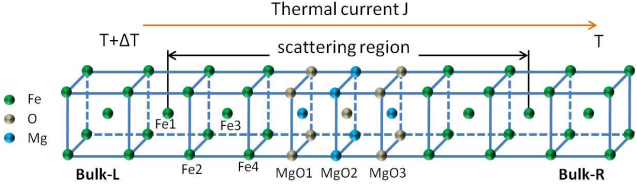}}
\caption{\label{fig:structure} Sketch of a Fe/MgO(3 MLs)/Fe-MTJs used for phonon transmission calculation. The scattering region for the supercell calculation is indicated.}
\end{figure}

The phonon transmission function and eventually the corresponding contribution to the thermal conductance is calculated by using the atomistic Green's function (AGF) method, where details are described in Ref.~\cite{Bachmanna:2012} and references therein. In order to calculate the phonon transport the junction is divided into three parts shown in Fig.~\ref{fig:structure}. The scattering region consists of two Fe/MgO interfaces sandwiched by two semi-infinite Fe leads. After constructing the Green's function $G(\varepsilon)$, the phonon transmission function $t_{p}(\varepsilon)$ and the thermal conductance $\kappa_{p}$ is calculated by~\cite{Bachmanna:2012}
\begin{equation}\label{eqn:Phtrans}
\begin{aligned}
 t_p(\varepsilon)&=\frac{1}{(2\pi)^{2}}{\int}_{2DBZ}Tr[{\Gamma_L(\varepsilon)}G(\varepsilon){\Gamma_R(\varepsilon)}G^\dagger(\varepsilon)]\mathrm{d}q_{\|} \\
\kappa_{p}&=\frac{1}{h}{\int} \varepsilon \: t_p(\varepsilon)\frac{\partial f_{p}(\varepsilon,T)}{\partial T}\mathrm{d}\varepsilon
\end{aligned}
\end{equation}
where ${\Gamma_L}(\varepsilon)$ and ${\Gamma_R}(\varepsilon)$ are the broadening matrices of the bulk Fe leads, $f_p(\varepsilon,T)$ is the occupation function, and $q_{\|}$ is the wave vector in the two dimensional Brillouin zone (2DBZ). The Green's function is calculated from the interatomic force constants (IFCs). Thereby, we construct the Green's function of the whole MTJ from the IFCs of bulk Fe and the IFCs of the scattering region.

\begin{figure}
{\includegraphics[width=8.6cm]{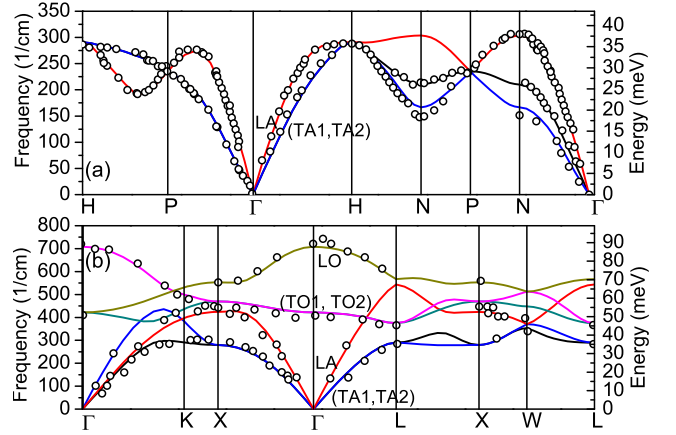}}
\caption{\label{fig:phononbulk} Phonon dispersions of (a) bcc-Fe and (b) fcc-MgO along high symmetry directions calculated by Abinit (color lines). The dotted data are experimental values taken from Refs.~\cite{Brockhouse:1967} and \cite{Sangster:1970}. Lines with different color correspond to different phonon branches.}
\end{figure}

In order to obtain the IFCs we use density functional perturbation theory (DFPT)~\cite{Gonze:1997} implemented in the Abinit package~\cite{Gonze:2005}. The self-consistent and perturbation calculations were performed by using Troullier-Martins norm-conserving pseudopotentials~\cite{Troullier:1991} and GTH LDA exchange-correlation functional~\cite{Goedecker:1996}.
First, to check the method we compute the phonon band structure for bulk Fe (a=2.867 {\AA}) and bulk MgO (a=4.239 {\AA}).
For the self-consistent calculation of bcc Fe (fcc MgO), we use an energy cutoff of 30 Ha (34 Ha) and a k-point mesh of $16\times16\times16$ ($8\times8\times8$). For the dynamic matrix we use for both materials a $8\times8\times8$ q mesh.
The calculated phonon dispersions are shown in Fig.~\ref{fig:phononbulk}. In addition, experimental values are shown for comparison and we find an excellent agreement.

To obtain the IFCs of the scattering region, a Fe/MgO/Fe supercell method is used. Thereby, the supercell consist of the MgO barrier with 4 monolayers of Fe at each side. The in-plane lattice constant is fixed to the value of bulk-Fe (a=2.867 {\AA}). The cell volume and atomic coordinates are fully relaxed until the forces on all atoms are smaller than 0.05 meV/{\AA}. After the structural relaxation, self-consistent and DFPT calculations are conducted according to different irreducible perturbations. For the self-consistent calculations we use a $10\times10\times2$ k point mesh and an energy cutoff of 34 Ha. For DFPT calculations, we use a $5\times5\times1$ q mesh with 6 inequivalent q points in the Brillouin zone. There are in total 120, 150, 180 and 210 independent DFPT calculations for 3, 5, 7, and 9 monolayers of MgO, respectively.

\begin{figure}
{\includegraphics[width=8.5cm]{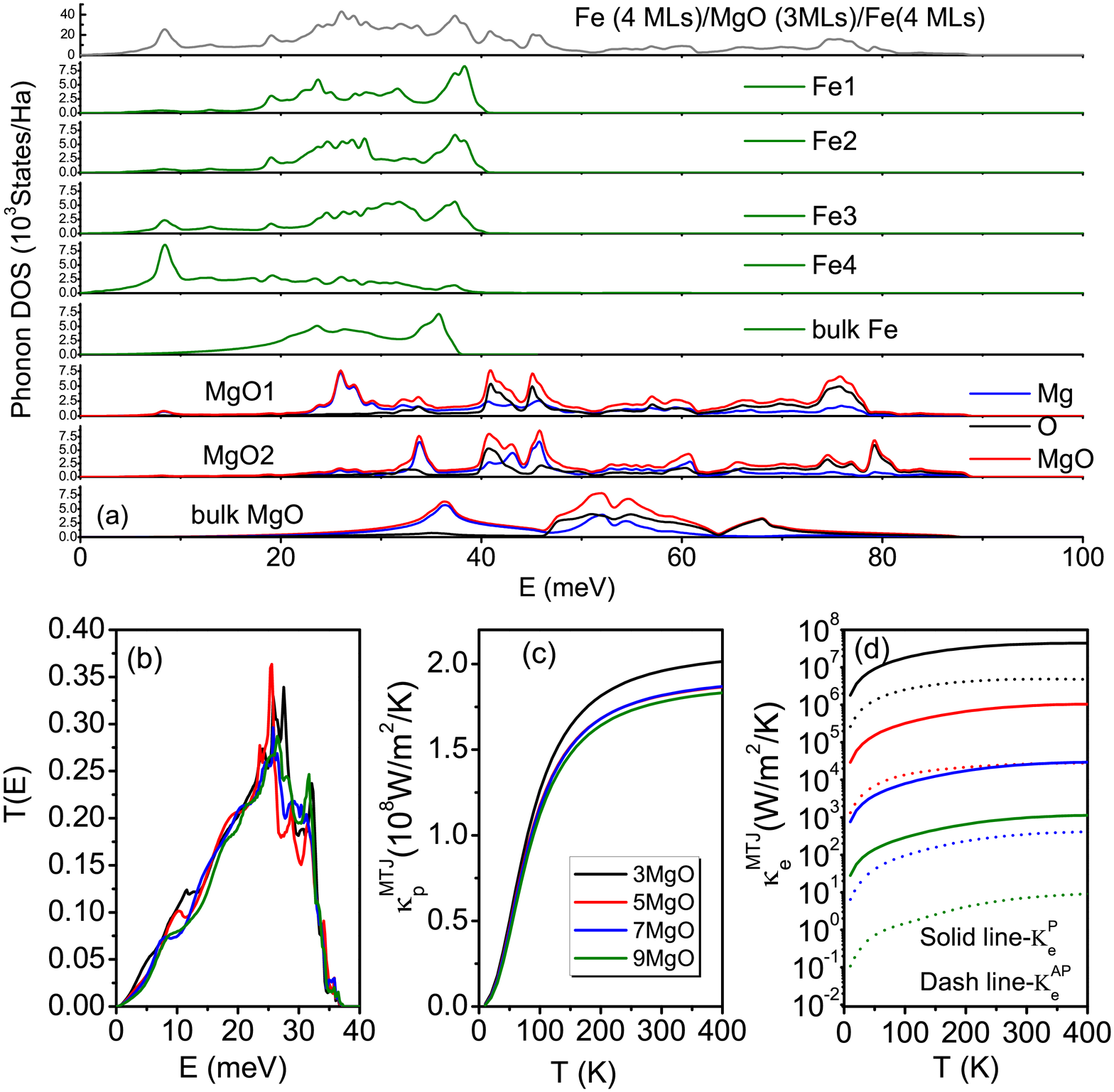}}
\caption{\label{fig:phtrans} (a) Projected phonon density of states (PDOS) for a supercell with 8 MLs Fe and 3 MLs MgO. The atom index is the same as in Fig.~\ref{fig:structure}. The PDOS for bulk Fe and MgO is shown for reference. (b)~The phonon transmission as a function of energy for Fe/MgO(3-9 MLs)/Fe-MTJs.(c) The phonon and (d) electron thermal conductance of Fe/MgO(3-9 MLs)/Fe-MTJs as a function of temperature.}
\end{figure}

The phonon density of states (PDOS) projected on each atom in the scattering region is shown in Fig.~\ref{fig:phtrans}(a).
The PDOS for the Fe layer near the MgO interface is significantly different from the bulk PDOS of Fe due to the interface phonon bonding modes between Fe-MgO. This pronounced interface phonon peak is located at 8.5 meV. Inside the MgO barrier there are several phonon peaks. However, compared with the bulk phonon modes of MgO, the peak positions inside the MgO barrier shift to different energies. The phonon modes at the Fe-MgO interface and inside the MgO barrier may take effect through the electron-phonon interaction and they can be observed by inelastic electron tunneling spectra (IETS)\cite{Drewello:2009}. For example, an IETS peak at 80 meV in $d^2I/dV^2$ curve has been attributed to the Mg-O surface phonon modes in MgO\cite{Drewello:2009} and this value is close to the largest peak position present in our projected PDOS.

After obtaining the dynamic matrix, the IFCs in the junction region are constructed. Away from the interface, the force constants are assumed to be that of bulk Fe. Our calculated phonon transmission function for different MgO thicknesses are shown in Fig.~\ref{fig:phtrans}(b). The cut-off energy of the transmission is at 38 meV and corresponds to the cut-off frequency of bulk-Fe.
The general shape of the transmission function is almost independent of the MgO thickness. Only some shifts of the peak positions are visible, which may come from quantized phonon modes within the MgO barrier. Using the phonon transmission, the phonon thermal conductance of the MTJs $\kappa_{p}^{MTJ}$ can be calculated by using eq.~(\ref{eqn:Phtrans}) and its temperature dependence is shown in Fig.~\ref{fig:phtrans}(c). There is a slight difference of $\kappa_{p}^{MTJ}$ between the different MgO thicknesses. At 300 K $\kappa_{p}^{MTJ}$ is found to be at the order of $10^{8} W/m^2/K$.

The screened-KKR Green's function method~\cite{Czerner:2011} is adopted to calculate the electronic transmission $t_{e}(\varepsilon)$ (details of the computation method are described in Refs.~\cite{Czerner:2011,Christian:2013}). $\kappa_{e}^{MTJ}$ is obtained from $t_{e}(\varepsilon)$ by~\cite{Liu:2009,Bachmann:2012b}
\begin{equation}
\kappa_{e}^{MTJ}=\frac{1}{T}[{L_2} - \frac{(L_1)^2}{L_0}]
\end{equation}
where $L_{n}=-\frac{2}{h}\int(\varepsilon-\mu)^{n}\frac{\partial f_{e}}{\partial T}t_{e}(\varepsilon)\mathrm{d} \varepsilon$, $\mu$ is the electro-chemical potential, and $f_{e}$ is the Fermi-Dirac distribution function.
$t_{e}(\varepsilon)$ depends on the relative magnetic orientation of the Fe leads to each other.

The calculated $\kappa_{e}^{MTJ}$ when the magnetization of the two Fe electrodes are in parallel or antiparallel alignment are shown in Fig.~\ref{fig:phtrans}(d). Since the electrons are tunneling through the MgO, $\kappa_{e}^{MTJ}$ is quite small and decreases exponentially with increasing barrier thickness. Even at the smallest considered MgO thickness of 3 monolayers $\kappa_{e}^{MTJ}$ at 300 K is 1 order of magnitude smaller than $\kappa_{p}^{MTJ}$.

After obtaining $\kappa_{p}^{MTJ}$ and $\kappa_e^{MTJ}$, the temperature profile in Fe/MgO/Fe-MTJs can be calculated by solving a one dimensional thermal transport equation with appropriate boundary conditions. We consider the electron and phonon transport, which carry the heat flux across the junction as well as the electron-phonon interaction in the Fe leads. It is reasonable to neglect the magnon contribution to the thermal transport since the magnon transport across the junction is assisted by the electron tunneling through electron-magnon interaction, which is at least one order of magnitude smaller than the electron and phonon transport process. The two fluid model which take into account the electron-phonon non-equilibrium in the Fe electrodes will result in different electron and phonon temperature $T_e(z)$ and $T_p(z)$ at the Fe-MgO interfaces.

The energy balance equations between electrons and phonons in the Fe leads are~\cite{Arun:2004}
\begin{equation}\label{eqn:kappa}
\begin{aligned}
k_e^{Fe}\frac{d^2T_e(z)}{dz^2}-G_{eph}(T_e-T_p)=0\\
k_p^{Fe}\frac{d^2T_p(z)}{dz^2}+G_{eph}(T_e-T_p)=0
\end{aligned}
\end{equation}
where $k_e^{Fe}$ and $k_p^{Fe}$ are the electron and phonon thermal conductivities in Fe, respectively. For these quantities we take the bulk values, which are listed in Table~\ref{tab:table1}. $G_{eph}$ is the electron-phonon interaction factor, which is determined by $G_{eph}=\pi{k_B}{\lambda}\langle\omega^2\rangle D(\varepsilon_F)$~\cite{Lin:2008}. In this expression, $k_B$ is the Boltzmann constant, $\lambda$ is the electron-phonon mass enhancement parameter, $D(\varepsilon_F)$ is the electron density of states at the Fermi energy, and $\langle\omega^2\rangle$ is the second moment of the phonon spectrum defined by McMillan~\cite{McMillan:1968}. $\langle\omega^2\rangle$ can be approximated by $\langle\omega^2\rangle\approx\theta_D^2/2$ where $\theta_D$ is the Debye temperature of Fe.  By using the related parameters listed in Table.~\ref{tab:table2} $G_{eph}$ of Fe is found to be $9.925\times10^{17} W/m^3/K$. This value for Fe is comparable but slightly smaller than that of Nickel ($10.5\times10^{17} W/m^3/K$) and Platinum ($10.9\times10^{17} W/m^3/K$)~\cite{Lin:2008}.
\begin{table}
\caption{\label{tab:table2}Parameters used for calculating the electron-phonon interaction factor of Fe.}
\centering
\begin{ruledtabular}
\begin{tabular}{ccc|cc}
  $\lambda$ & $\theta_D$ (K)& $D(\varepsilon_F)$ (States/Ha)& $G_{eph}$ ($W/m^3/K$)\\
  \hline
  0.243~\cite{Matthieu:2013} & 470~\cite{Kittel:1986} & 23.02~\cite{Matthieu:2013} & $9.925\times10^{17}$ \\
\end{tabular}
\end{ruledtabular}
\end{table}

The solutions of Eq. (\ref{eqn:kappa}) for the left (L) and right (R) Fe layers are
\begin{equation} \label{eqn:sol}
  T_{e,p}^{L(R)}(z)=B_{e,p}^{L(R)} + C_{e,p}^{L(R)} z + D_{e,p}^{L(R)} e^{+(-) \frac{z}{l_{Fe}}}
\end{equation}
where the coefficients of the electrons and phonons hold
\begin{equation}\label{eqn:coeff} \nonumber
B_e^{L(R)}=B_p^{L(R)} \ ; \ C_e^{L(R)}=C_p^{L(R)} \ ; \ D_e^{L(R)}=- \frac{k_p^{Fe}}{k_e^{Fe}}D_p^{L(R)}
\end{equation}
Thus the last term in Eq. (\ref{eqn:sol}) accounts for the electron-phonon imbalance at the Fe/MgO interface. Consequently,
\begin{equation}\label{eqn:l}
l_{Fe}= \sqrt{\frac{1}{G_{eph}}\frac{k_e^{Fe} k_p^{Fe}}{k_p^{Fe}+k_e^{Fe}}}
\end{equation}
is the characteristic length of this imbalance. We define
\begin{equation}\label{eqn:l2}
l_{e,p}=\sqrt{\frac{\kappa_{e,p}^{Fe}}{G_{eph}}} \ \ \Rightarrow \ \ 1/{l_{Fe}^2}=1/l_e^2+1/l_p^2;
\end{equation}
In total we have to calculate 6 coefficients $B_e^{L(R)}$, $C_e^{L(R)}$, and $D_e^{L(R)}$ by using the following 6 boundary conditions
\begin{eqnarray*}
\nonumber \frac{dT_{e,ph}^{L}(z)}{dz}\mid_{z=-d/2}&=&\frac{dT_{e,ph}^{R}(z)}{dz}\mid_{z=d/2} \\
\nonumber \kappa_{e,ph}^{MTJ} \left (T_{e,ph}^L(\frac{-d}{2})-T_{e,ph}^R(\frac{d}{2}) \right )&=&-\kappa_{e,ph}^{Fe} \frac{dT_{e,ph}^{R}(z)}{dz}\mid_{z=d/2} \\
T_e^L(z=-L)=T_L \ &\ &\ \ \ T_e^R(z=L)=T_R
\end{eqnarray*}
where the last two conditions are some given temperatures at left and right of the MTJ, which has a total thickness of $2 L$.



\begin{figure}
{\includegraphics[width=8.7cm]{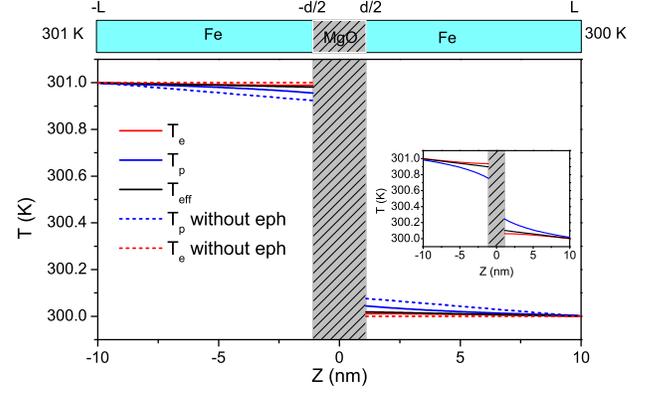}}
\caption{\label{fig:TMTJ} The electron temperature $T_e$, the phonon temperature $T_p$ and the linear extrapolation of electron temperature $T_{eff}$ profiles across a Fe/MgO(9 MLs)/Fe-MTJ with the temperature at the left side (L=-10 nm) $T_L=301 K$ and right side (L=10 nm) $T_R=300 K$. The inset shows the temperature profiles assuming a ten times larger $\kappa_p^{MTJ}$ than the first principle value.}
\end{figure}

To illustrate our result we show in Fig.~\ref{fig:TMTJ} the electron and phonon temperature profiles for a Fe(10 nm)/MgO(9 MLs)/Fe(10 nm)-MTJ with a temperature difference of 1 K.
The consequence of different electron and phonon temperatures near the MgO/Fe interface is significant especially for the definition and evaluation of thermoelectric physical quantities. For example, the Seebeck coefficient S is defined as $S=\frac{{\Delta}V}{{\Delta}T_e}$ where ${\Delta}T_e$ is the electron temperature drop since the Seebeck effect is a consequence of electron transport.

Although we have a non-equilibrium situation it might be useful to define an effective temperature drop $\Delta T_{eff}$, which can be used to define a $\kappa^{MTJ}=\frac{q}{{\Delta}T_{eff}}$, where $q=-k_e^{Fe} \frac{dT_e^{R}(z)}{dz}-k_p^{Fe} \frac{dT_p^{R}(z)}{dz}$ is the total thermal current through the junction. Note that $q$ is conserved whereas the electron and phonon thermal current alone are not due to the imbalance.

Following Ref. \cite{Arun:2004} we define the effective temperature $T_{eff}$ by linear extrapolation of the electron or phonon temperature from the equilibrium towards the MgO barrier (see Fig.~\ref{fig:TMTJ}).
For the case shown in Fig.~\ref{fig:TMTJ} we calculate $\kappa^{MTJ}=1.681\times10^8 W/m^2/K$. This value can be used in a simple network model to estimate the electron temperature drop by calculating $\Delta T_{eff}$. To get an idea about the error in this estimation we calculate the following ratio
\begin{equation}
\frac{T_e-T_{eff}}{T_{eff}-T_p}=\frac{k_p^{Fe}}{k_e^{Fe}}
\end{equation}
This means if $k_e^{Fe} \gg k_p^{Fe}$, which is the case for most metals, $T_{eff}$ is closer to the electron temperature than to the phonon temperature. If this is not the case or a more precise knowledge of $\Delta T_e$ is needed one has to solve appropriate transport equations, i.e. Eq. (\ref{eqn:kappa}), for the whole junction, but taking the first principle values for $\kappa_e^{MTJ}$ and $\kappa_p^{MTJ}$.
Moreover, the effect of the imbalance is larger if the phonon interface conductance is larger. This effect is shown in the inset of Fig.~\ref{fig:TMTJ} where we assume a ten times larger $\kappa_p^{MTJ}$ than the first principle value, which can be achieved by using different materials. Note that even the phonon temperature drop across the barrier decreases the drop in the electron temperature remains.


\begin{figure}
{\includegraphics[width=8.7cm]{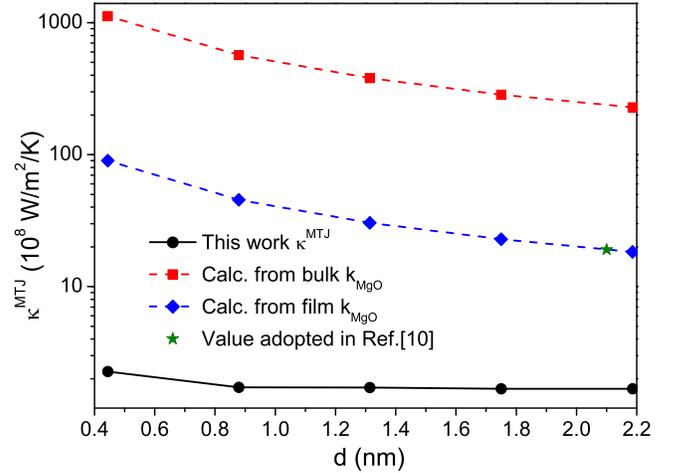}}
\caption{\label{fig:kappa} The junction thermal conductance as a function of MgO barrier thickness in this work (black) compared to bulk MgO (red dashed) and thin film MgO (blue dashed). The value used by Walter \textit{et.al.} \cite{Walter:2011} is shown as a green star.}
\end{figure}

The thermal conductance $\kappa^{MTJ}$ of a MTJ is of particular importance for the interpretation of experimental results. In experiments a Seebeck voltage is measured and a temperature drop is estimated using diffusion models. As stated earlier often a value of bulk or thin film value of MgO is used for $\kappa^{MTJ}$. In Fig.~\ref{fig:kappa} we plot our calculated $\kappa^{MTJ}$ in comparison to bulk and thin film values of MgO for different MgO thicknesses.
Our results are almost independent of the MgO thickness. Further, as an example the value used in Ref.~\cite{Walter:2011} is indicated by the green star. This implies that the estimated Seebeck coefficients in Ref.~\cite{Walter:2011} may be too high.


In conclusion, we show that there is an imbalance of the electron and phonon temperature at nano-magnetic interfaces. This leads to different temperature drops for electrons and phonons and one has to be cautious interpreting results. Nevertheless, in the considered Fe/MgO/Fe MTJ we observed a large interface resistance for the phonons. In consequence, a large drop also for the electron temperature exists, which is responsible for high Seebeck voltages observed in experiments. But even for MTJs with low phonon interface resistances the drop of the electron temperature remains large due to the strong imbalance of electron and phonon temperature. For a MTJ with metallic leads we define an effective thermal conductance by defining an effective temperature. The corresponding values given in Fig.~\ref{fig:kappa} can be used in simple models as the conductance of the MTJ (barrier plus interfaces). However, if more reliable values of the temperature drops are needed a calculation of the coupled electron and phonon system is necessary. In any case for the qualitative description first principle calculations are a need due to the coherent transport across the nano-magnetic interface.


We thank Prof. M. Munzenberg and Prof. Matthieu Verstraete for useful discussions and acknowledge support from Deutsche Forschungsgemeinschaft (SPP 1538 via Grant No. HE 5922/4-2). We acknowledge support within the LOEWE program of excellence of the Federal State of Hessen (project initiative STORE-E). This project was supported by the Laboratory of Materials Research (LaMa) of JLU.


\begin{thebibliography}{99}
\bibitem{Johnson:1987} M. Johnson, R.H. Silsbee, Phys. Rev. B {\bf 35}, 4959 (1987).
\bibitem{Bauer:2010} G. E. W. Bauer, A. H. MacDonald, and S. Maekawac, Solid State
Commun. {\bf 150}, 459 (2010)
\bibitem{Bauer:2012}G. E. W. Bauer, E. Saitoh, and B. J. van Wees, Nat. Mater. {\bf 11}, 391 (2012).
\bibitem{Uchida:2008} K. Uchida, S. Takahashi, K. Harii, J. Ieda, W. Koshibae, K. Ando, S. Maekawa, and E. Saitoh, Nature (London) {\bf 455}, 778 (2008).
\bibitem{Xiao:2010} J. Xiao, G. E. W. Bauer, K. Uchida, E. Saitoh, and S. Maekawa, Phys. Rev. B {\bf 81}, 214418 (2010)
\bibitem{Breton:2011} J. C. Le Breton, S. Sharma, H. Saito, S. Yuasa, and R. Jansen, Nature (London) {\bf 475}, 82 (2011).
\bibitem{Hatami:2009} M. Hatami, G. E. W. Bauer, Q. F. Zhang, and P. J. Kelly, Phys. Rev.
B {\bf 79}, 174426 (2009).
\bibitem{Flipse:2012} J. Flipse, F. L. Bakker, A. Slachter,	F. K. Dejene and B. J. van Wees, Nature Nanotech. {\bf 7}, 166–168 (2012).
\bibitem{Flipse:2014} J. Flipse, F. K. Dejene, D. Wagenaar, G. E. W. Bauer, J. Ben Youssef, and B. J. van Wees, Phys. Rev. Lett. {\bf 113}, 027601 (2014)
\bibitem{Walter:2011}  M. Walter, \textit{et.al.}, Nat. Mater. {\bf 10}, 742 (2011).
\bibitem{Liebing:2011} N. Liebing, S. Serrano-Guisan, K. Rott, G. Reiss, J. Langer, B. Ocker, and H. W. Schumacher, Phys. Rev. Lett. {\bf 107}, 177201 (2011)
\bibitem{Lin:2012} W. Lin, M. Hehn, L. Chaput, B. Negulescu, S. Andrieu, F. Montaigne, and S. Mangin, Nature Commun. {\bf 3}, 744 (2012).
\bibitem{Czerner:2011} M. Czerner, M. Bachmann, and C. Heiliger, Phys. Rev. B {\bf 83}, 132405 (2011)
\bibitem{Christian:2013} C. Heiliger, C. Franz, and M. Czerner, Phys. Rev. B {\bf 87}, 224412 (2013)
\bibitem{Jia:2011} X.-T. Jia, K. Xia and G. E. W. Bauer, Phys. Rev. Lett. {\bf 107}, 176603 (2011)
\bibitem{Christian:2014} C. Heiliger, C. Franz, and M. Czerner, J. Appl. Phys. {\bf 115}, 172614 (2014)
\bibitem{leut:2013} J. C. Leutenantsmeyer, M. Walter, V. Zbarsky, M. Münzenberg, R. Gareev, K. Rott, A. Thomas, G. Reiss, P. Peretzki, H. Schuhmann, M. Seibt, M. Czerner, C. Heiliger, SPIN {\bf 03}, 1350002 (2013)
\bibitem{Chen:1998} G. Chen, Phys. Rev. B {\bf 57}, 14958 (1998)
\bibitem{Lee:1998} S.-M. Lee, David G.Cahill, and Thomas H. Allen, Phys. Rev. B {\bf 52}, 253 (1995)
\bibitem{Williams:1981} R. K. Williams, D. W. Yarbrough, J. W. Masey, T. K. Holder, and R. S. Graves, J.Appl.Phys.{\bf 52},5167 (1981)
\bibitem{Anne:2014} A. M. Hofmeister, Phys. Chem. Minerals. {\bf 41}, 361(2014)
\bibitem{Glen:1962} G. A. Slack, Phys. Rev. {\bf 126}, 427(1962)
\bibitem{Karki:2000} B. B. Karki, R. M. Wentzcovitch, S. de Gironcoli and S. Baroni, Phys. Rev. B {\bf 61}, 8793 (2000).
\bibitem{Bachmanna:2012} M. Bachmanna, M. Czerner, S. Edalati-Boostan, and C. Heiliger, Eur. Phys. J. B  {\bf 85}, 146(2012)
\bibitem{Gonze:1997} X. Gonze, Phys. Rev. B {\bf 55}, 10337 (1997); X. Gonze, C. Lee, Phys. Rev. B {\bf 55}, 10355 (1997)
\bibitem{Gonze:2005} X. Gonze, G.-M. Rignanese, M. Verstrate, J.-M. Beuken, Y. Pouillon, R. Caracas, F. Jollet, M. Torrent, G.Zerah, M. Mikami, P. Ghosez, M. Veithen, J.-Y. Raty, V. Olevano, F. Bruneval, L. Reining, R. Godby, G. Onida, D.R. Hamann, D.C. Allan, Z. Kristallogr. {\bf 220}, 558 (2005)
\bibitem{Troullier:1991} N. Troullier, J.L. Martins, Phys. Rev. B {\bf 43}, 1993 (1991)
\bibitem{Goedecker:1996} S. Goedecker, M. Teter, J. Huetter, Phys. Rev. B {\bf 54}, 1703 (1996)
\bibitem{Brockhouse:1967} B.N. Brockhouse, H.E. Abou-Helal, and E.D. Hallman, Solid State Commun. {\bf 5}, 211 (1967).
\bibitem{Sangster:1970} M. J. L. Sangster, G. Peckham, and D. H. Saunderson, J. Phys. C. {\bf 3}, 1026 (1970).
\bibitem{Drewello:2009} V. Drewello, M. Sch\"{a}fers, O. Schebaum, A. A. Khan, J. M\"{u}nchenberger, J. Schmalhorst, G. Reiss, and A. Thomas, Phys. Rev. B 79, 174417 (2009).
\bibitem{Liu:2009} Yu-Shen Liu, Yi-Ren Chen, and Yu-Chang Chen, ACS Nano {\bf 3}, 3497 (2009).
\bibitem{Bachmann:2012b} M. Bachmann, M. Czerner, and C. Heiliger, Phys. Rev. B {\bf 86}, 115320 (2012).
\bibitem{Arun:2004} A. Majumdara, P. Reddy, Appl. Phys. Letts. {\bf 84}, 4768 (2004)
\bibitem{Lin:2008} Z. Lin, L. V. Zhigilei, and V. Celli, Phys. Rev. B {\bf 77}, 075133 (2008).
\bibitem{McMillan:1968} W. L. McMillan, Phys. Rev. {\bf 167}, 331 (1968).
\bibitem{Matthieu:2013} M. J. Verstraete, J. Phys.: Condens. Matter {\bf 25}, 136001 (2013).
\bibitem{Kittel:1986} C. Kittel, Introduction to Solid State Physics, 6th ed. (Wiley, New York, 1986).
\end{thebibliography}
\end{document}